\documentclass[a4paper,11pt]{article}
\usepackage{pos}

\title{Neutrino theory: open questions and future opportunities}
\author{Raymond R. Volkas}
\affiliation{ARC Centre of Excellence for Dark Matter Particle Physics,\\
  School of Physics, The University of Melbourne, Victoria 3010, Australia}

\emailAdd{raymondv@unimelb.edu.au}
\abstract{The subtitle of my talk is ``The quest for understanding the origin of neutrino masses''. After reviewing why the discovery of neutrino masses is also the discovery of New Physics, the substance of the talk details mechanisms for generating Majorana neutrino masses and implications for experimental searches and/or cosmology. I review high-scale seesaw, low-scale seesaw and radiative mechanisms, asking at every turn how testable the scenario is. While it is clear that determining the origin of neutrino masses -- knowing what Lagrangian to put into textbooks -- is a distant and ambitious goal, I end with experimental advances that we can reasonably hope for that would constitute progress.}

\FullConference{%
  42nd International Conference on High Energy Physics\\
  17-24 July 2024\\
  Prague, Czech Republic
}

\begin{document}
\maketitle

\section{Introduction}

What exactly is a quest? According to the online Britannica Dictionary a quest can mean ``a long and difficult effort to find or do something'' as in the phrase ``a quest for answers'' and the sentence ``They refuse to give up their quest to discover the truth''. In the worlds of myth and fiction, famous quests include Sir Galahad seeking the Holy Grail -- which may be a cup, or maybe not -- and Don Quixote's mad quest to restore chivalry. Our quest is to find out how to extend the standard model (SM) Lagrangian to include the neutrino mass and mixing mechanism actually used by Nature. We have invented a plethora of possibilities, but are missing the experimental results needed to discard the pretenders.

Are we mad like Don Quixote? Is our fate merely to establish some facts about neutrino masses and mixings, but fall short of uncovering the Lagrangian? It may well be, but, like Sir Galahad, we can live in hope. Our Grail quest may provide some really important information about the fundamental mechanism, even if the full picture proves unobtainable.

The main goal of this talk is to survey three of the principal neutrino mass generation paradigms -- high-scale seesaw, low-scale seesaw and radiative -- and comment on their experimental testability and other implications.\footnote{Apologies for topics I will not have the time to cover. They include flavour symmetry models, oscillation parameter fitting, neutrino interactions, CE$\nu$NS or coherent neutrino scattering, sterile neutrinos, astrophysical signals, and cosmological bounds. Please consult the many excellent talks on these topics at both this conference and the recent Neutrino 2024 meeting.} But first, one short digression and two preambles.

\section{Establishing the SM charged-fermion mass generation mechanism}

I phrased our quest above as a long and difficult effort to find out how to extend the SM Lagrangian. But this supposes that the SM Lagrangian has in fact been established to good accuracy, especially in the charged-fermion mass generation sector.\footnote{Of course, other terms in the Lagrangian, such as the Higgs boson self-coupling, also need to be experimentally confirmed. I highlight the Yukawa sector because it pertains to fermion mass generation.} This, clearly, is not the case. We have no experimental information about how strongly the physical Higgs boson couples to the light quarks and the electron. And, it might surprise you to learn that 1-loop alternatives to the SM tree-level mechanism for generating the relatively large $b$ and $\tau$ masses are phenomenologically viable~\cite{Baker:2020vkh,Baker:2021yli}. Current experimental results are consistent with the SM mechanism, but not precise enough to rule out this alternative. Indeed, such an alternative remains theoretically well motivated because of the observed hierarchical pattern of charged-fermion masses. 

\section{$\nu$ mass is New Physics}

In the minimal SM, neutrinos are massless, and the family lepton numbers $L_{e,\mu,\tau}$ are perturbatively conserved, because of the absence of both right-handed (RH) neutrinos and hypercharge $Y=1$ scalar triplets. When made massive, neutrinos are either Dirac or Majorana fermions. If they are Majorana fermions, then they are the first such elementary particles discovered, and thus constitute New Physics (NP). There are a very large number of viable mechanisms for generating Majorana masses, and they \textit{all} involve new degrees of freedom. If, on the other hand, neutrinos are Dirac fermions, then RH neutrinos are needed. They are new degrees of freedom and hence NP. Also, in the context of the SM gauge group, the renormalisable and gauge-invariant Majorana mass terms for the RH neutrinos have to be expelled from the Lagrangian, which constitutes a new model-building principle: the imposition of global total lepton number symmetry. No matter what, $\nu$ mass requires NP.

\section{Is $\nu$ mass a new physics scale?}

While we have yet to measure the absolute neutrino mass scale, we know that it is tiny. The KATRIN beta decay endpoint measurements now provide an upper bound of $0.45$ eV on this mass scale, which is an impressive achievement~\cite{Katrin:2024tvg}. An even stronger bound is obtained from large-scale structure and CMB considerations in cosmology, assuming the standard $\Lambda$CDM model is correct. The precise figure depends on what data sets are combined and how some systematics are handled -- see, for example, Ref.~\cite{Allali:2024aiv} for a critical discussion. But an upper bound of order $0.1$ eV seems to be indicated. So, the neutrino mass scale is very small compared to all other known scales, even the electron mass, except for that of dark energy.

But now we should ask if the \textit{origin} of the tiny neutrino mass scale relies on a genuinely new physical mass scale, or is instead derived from a known scale (or known scales) via tiny dimensionless parameters? For example, if neutrino masses are generated in exactly the same way as the charged fermions in the SM, then they are equal to the Higgs vacuum expectation value (VEV) multiplied by tiny Yukawa coupling constants of order $10^{-12}$ or less.

There is nothing technically wrong with this. And, we know that the SM requires some small dimensionless parameters: the electron Yukawa coupling constant is of order $10^{-6}$, for example. If $10^{-6}$ is OK, is $10^{-12}$ also OK? Technically, as I said, it is fine, but to most people it suggests that some more intricate physics is involved. 

For the remainder of this talk I assume that the $\nu$ mass scale \textit{does} imply a new physical scale, and also that neutrino masses are Majorana. Model building strategies have been dominated by providing a rationale for the tiny neutrino mass scale and proposing various ways to break lepton number conservation.

\section{Tree-level high-scale seesaw models}

It remains an intriguing fact that the lowest mass dimension non-renormalisable effective operators formed from gauge-invariant products of SM fields are of the Weinberg form~\cite{Weinberg:1979sa},
\begin{equation}
\frac{\lambda_{ij}}{M} L_i L_j H H,
\end{equation}
where the $L_i$ are LH lepton doublets, $H$ is the Higgs doublet, $M$ is an NP scale and $\lambda_{ij}$ are dimensionless coupling constants. These operators explicitly break lepton number by two units. Replacing the Higgs fields with VEVs, we thus immediately obtain a Majorana mass matrix of the famous seesaw form,
\begin{equation}
    m_{\nu,ij} = \lambda_{ij} \frac{v^2}{M},
\end{equation}
where $v \simeq 175$ GeV is the VEV. By requiring that $M \gg v$, we get that $m_\nu \ll v$ and thus have a very simple explanation for the tiny neutrino mass scale. Using $0.1$ eV for that scale, we obtain $M \sim 10^{14}$ GeV for the NP scale assuming that the $\lambda \sim 1$. This is a very high value, and it makes the minimal, pure seesaw scenario essentially impossible to test.

Before commenting on this point further, let us figure out UV completions for the Weinberg operators. Famously, there are three minimal, tree-level possibilities: the type 1, 2 and 3 seesaw models, which differ by the nature of the mediator that is integrated out to produce the effective operators. One possibility sees $LH$ coupling to new fermions, which may be SM gauge singlets (i.e.\ RH neutrinos) giving the type 1 seesaw model~\cite{Minkowski:1977sc,Yanagida:1979as,Glashow:1979nm,GellMann:1980vs,Mohapatra:1979ia}, or isospin triplet, hypercharge zero multiplets giving the type 3 seesaw model~\cite{Foot:1988aq}. The other possibility is both $LL$ and $HH$ couple to a scalar isospin triplet with unit hypercharge, giving the type 2 seesaw model~\cite{Konetschny:1977bn,Magg:1980ut,Schechter:1980gr,Cheng:1980qt,Mohapatra:1980yp,Wetterich:1981bx}. In each case, the NP scale $M$ is given by the mass scale of the mediator(s). Lepton number violation is driven by Majorana mass terms for the mediators in the type 1 and 3 cases. In the type 2 case, the scalar triplet acquires $L=2$ from coupling to $LL$ and a trilinear coupling between it and an $L=0$ Higgs doublet bilinear breaks lepton number by two units.

Note that the process of starting with effective operators and then deriving all the minimal UV completions has the advantage that no possibilities are missed. (Historically, this did not happen: the type 3 seesaw model was invented a decade after the related type 1 model.) This model-building strategy will be used when we turn to radiative models below.

We now discuss the pros and cons of the high-scale seesaw models. 

The type 1 model is the ``obvious'' way to generate neutrino masses via a minimal extension of the SM. You simply add two or three gauge-singlet RH neutrinos to the minimal SM particle content and write down the most general gauge-invariant renormalisable Lagrangian. That is a pro. We define the minimal, pure type 1 seesaw model through
\begin{itemize}
    \item $M \gg v$,
    \item all $\bar{L}\tilde{H}\nu_R$ Yukawa coupling constants are of order one or at least not tiny,
    \item the gauge group is that of the SM,
    \item the RH neutrinos are gauge singlets,
    \item the RH Majorana masses are bare masses, not generated by a VEV.
\end{itemize}
As stated above, the lack of testability of this pure incarnation is a con. But it also permits the very elegant and minimal baryogenesis via leptogenesis scenario whereby the cosmological matter-antimatter asymmetry is generated by CP-violating out-of-equilibrium decays of the heavy RH Majorana neutrinos (also called ``heavy neutral leptons (HNLs)'') into lepton and Higgs doublets and their CP conjugates~\cite{Fukugita:1986hr}. This is another pro. 

One way to depart from the pure realisation is to take the $\bar{L}\tilde{H}\nu_R$ Yukawa coupling constants to be much less than one, so that $M$ has to be reduced. Because the RH neutrinos are gauge sterile, and the Yukawa couplings are now small, experimental testability is still challenging. But an important example shows that there are prospects. The $\nu$MSM~\cite{Asaka:2005an,Asaka:2005pn} takes the lightest of the RH neutrinos to have a few-keV scale mass and thus is a warm dark matter candidate. The two other RH neutrinos are more massive and the dynamics is such that baryogenesis via a different type of leptogenesis is possible. This model motivates the search for heavy neutral leptons in meson decays (using, for example, the SHiP detector~\cite{Alekhin:2015byh}). In general, the search for HNLs at all practicable energy scales is important because it is a direct probe of neutrino mass generation~(see Ref.~\cite{Drewes:2024bla} for a recent analysis).\footnote{In another class of extensions the RH Majorana masses are associated with a cosmological symmetry-breaking phase transition that produces defects such as cosmic strings and can generate a gravitational wave signature. See, for example, Ref.~\cite{Fornal:2024avx} for a recent analysis. In particular, the seesaw scale may be identified with the Peccei-Quinn scale as in $\nu$DFSZ/VISH$\nu$~\cite{Clarke:2015bea,Sopov:2022bog} and SMASH~\cite{Salvio:2015cja,Ballesteros:2016euj}.}

The type 2 and 3 seesaw models have better prospects for discovery in high-energy colliders because the exotics couple to electroweak bosons. However, that still requires the seesaw effect to be somewhat compromised to bring the scale $M$ down to the TeV level.

While the type 1 model is ``obvious'' as a minimal SM extension, the situation is more complicated in gauge extensions where the RH neutrino is embedded in a multiplet of a non-Abelian group. For example, in the most common version of the left-right symmetric model (LRSM) a scalar triplet VEV is used to generate RH Majorana masses, while in SO(10) grand unified theories a 126-dimensional multiplet is employed. The connection with a higher symmetry breaking scale may be seen as attractive, but if you want to avoid higher-dimensional multiplets, there are alternatives that we now discuss that come with additional benefits.

\section{Low-scale tree-level seesaw models}

The most studied models in the class are the inverse seesaw (ISS)~\cite{Mohapatra:1986aw,Mohapatra:1986bd,Ma:1987zm} and linear seesaw (LSS)~\cite{Akhmedov:1995vm,Barr:2003nn,Malinsky:2005bi} models. It is useful to frame their review in the context of SO(10).\footnote{Analogous comments can be made for the Pati-Salam and LRSM subgroups.} Each family of fermions is placed in the $16$ of SO(10) and gauge sterile fermions $S_R$ are added. The general neutrino Yukawa matrix with family indices suppressed is
\begin{equation}
    \left( \begin{array}{ccc} \overline{\nu}_L & \overline{(\nu_R)^c} & \overline{(S_R)^c} \end{array} \right) \left( \begin{array}{ccc} 126 & 10+120 & 16 \\ 10+120 & 126 & 16 \\ 16 & 16 & 1 \end{array} \right) \left( \begin{array}{c} (\nu_L)^c \\ \nu_R \\ S_R  \end{array} \right)
\end{equation}
where the entries specify which scalar multiplets can contribute. The 120 and 126 are not necessary; the lower-dimensional 10 and 16 suffice to generate neutrino masses. Replacing those multiplets with VEVs results in the neutrino mass matrix
\begin{equation}
    \left( \begin{array}{ccc}
       0  & m & m_L \\
       m  & 0 & m_R \\
       m_L & m_R & \mu
    \end{array} \right)
\end{equation}
where $\mu$ is a $\Delta L = 2$ bare Majorana mass matrix for the sterile fermions. The entry labelled $m$ is the ``usual'' neutrino Dirac mass connecting $\nu_L$ and $\nu_R$, $m_L$ is an SU(2)$_L$ breaking Dirac-type mass linking $\nu_L$ with $S_R$, and $m_R$ is a higher-scale associated with the breaking of the SU(2)$_R$ subgroup.

The ISS has $m_L = 0$ and $\mu \ll m \ll m_R$ with the small value for $\mu$ being technically natural because total lepton number invariance is recovered in the $\mu =0$ limit. The light neutrino mass scale is given by
\begin{equation}
    m_\nu \sim \mu \left( \frac{m}{m_R} \right)^2
\end{equation}
so there is a triple suppression. This permits the scale of NP to be plausibly much lower than in the pure type 1 case. For example, setting $m_\nu \sim 0.1$ eV and $m_R \sim 10$ TeV requires $\mu/\text{MeV} \sim 10/(m/\text{GeV})^2$ so is not ridiculously small for GeV-scale $m$. A related advantage is that the active-sterile mixing angle is of order $m/m_R$, so relatively large compared to pure type 1. The LSS has $\mu = 0$ and $m_L \le m \ll m_R$, and the light neutrino mass scale is of order $m m_L/m_R$. Choosing the technically natural hierarchy $m_L \ll m$ results in double suppression. The scale of NP can again be fairly low without violating the spirit of the suppression mechanism. The ISS and LSS produce small Majorana neutrino masses with a relatively low NP scale without needing a small neutrino Dirac mass scale, but at the expense of smallish but technically-natural explicit $L$-violation scales given by $\mu$ and $m_L$. These scenarios are generically more testable than the minimal, pure type 1 seesaw.

\section{Radiative or loop-level models}

\begin{figure}[t]
\centering
\includegraphics[width=1.0\linewidth]{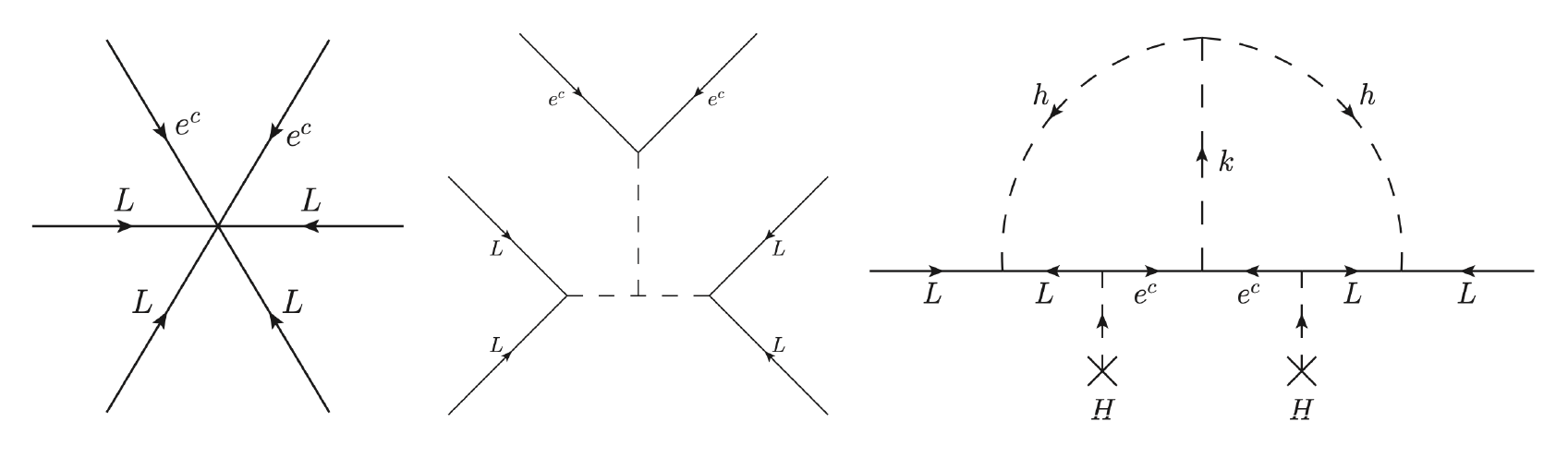}
\caption{The left panel shows the $\Delta L = 2$ effective operator $LLLe^cLe^c$. The middle panel shows one way it can be opened up at tree-level using singly- and doubly-charged scalars. The right panel shows how the external lines can be closed off to produce a Majorana neutrino mass 2-loop diagram (Zee-Babu model).
}
\label{f:ZeeBabu}
\end{figure}

There are more than $10^4$ such models! (See Ref.~\cite{Cai:2017jrq} for a comprehensive review of some of them.) Thus a systematic approach is called for. This is provided by starting with $\Delta L = 2$ effective operators and then opening them up to produce loop-level generation of Majorana neutrino masses. Two complementary approaches have been developed. One begins with generalised Weinberg operators of the form $LLHH(H^\dagger H)^n$ where $n=0,1,2,\ldots$. This strategy has been pursued by the Valencia group and collaborators~\cite{Cepedello:2017eqf,Cepedello:2018rfh,Arbelaez:2022ejo}. The other, pursued in recent years by the Melbourne group, considers non-Weinberg operators~\cite{Babu:2001ex,deGouvea:2007qla,Angel:2012ug,Cai:2014kra,Gargalionis:2020xvt} that are opened up at tree-level and then some external legs are closed off to produce neutrino self-energy diagrams. See Fig.~\ref{f:ZeeBabu} for how the historically important Zee-Babu model~\cite{Zee:1985id,Babu:1988ki} can be derived this way.

Radiative models explain the smallness of neutrino masses through the joint action of $1/16\pi^2$ suppression per loop, the relatively high masses of the exotics, and a product of coupling constants each of which may be somewhat less than one. While the exotics need to be massive enough to evade current constraints, the existence of the other two suppression factors means they can have masses as low as the TeV-scale. That means they can be directly searched for at colliders and their indirect or virtual effects may be detectable. Generic phenomenological implications include charged-lepton flavour violation, quark flavour effects, and contributions to magnetic and electric dipole moments. ``Scotogenic'' models feature dark matter candidates~\cite{Ma:2006km}.

Since there are so many radiative models, it makes sense to statistically survey their general features after a comprehensive database of systematically constructed models has been produced. Of course, some restrictions on the model-building algorithm must be imposed to render model construction a finite task. This kind of analysis was reported in Ref.~\cite{Gargalionis:2020xvt} for models adhering to the following criteria: (a) the gauge group is that of the SM, (b) the exotics are scalars, vectorlike Dirac fermions and Majorana fermions, and (c) $\Delta L = 2$ effective operators beyond mass dimension 11 are not considered. 

The computational analysis revealed 11,216 distinct models in this class. The most common scalar exotics are leptoquarks, with $R_2 \sim (3,2)(7/6)$ featuring most frequently, followed in decreasing order of frequency by $S_3 \sim (\bar{3},3)(1/3)$, $\tilde{R}_2 \sim (3,2)(1/6)$, $S_1 \sim (\bar{3},1)(1/3)$ and $\tilde{S}_1 \sim (\bar{3},1)(4/3)$. The next most common scalars are diquarks, followed by dileptons, with the remainder being outside of these categories because the given exotic directly couples to at least one other exotic. The most numerous exotic fermions are vectorlike quarks, followed by vectorlike leptons, and then other species. Many of these exotics have been independently motivated by anomalies in the flavour sector, including the now defunct $R_{K^{(*)}}$ anomalies, the still existing $R_{D^{(*)}}$ discrepancies, and $(g-2)_{\mu, e}$.\footnote{Space constraints permit only an incomplete list of relevant papers; see, for example, Refs.~\cite{Bauer:2015knc,Pas:2015hca,Becirevic:2016oho,Cai:2017wry,Dorsner:2017ufx,Buttazzo:2017ixm,Angelescu:2018tyl,Hati:2018fzc,Bigaran:2019bqv,Bigaran:2020jil}.} Whether the remaining anomalies are confirmed or disappear, the point is that the exotics in radiative neutrino mass models generically affect flavour observables in both the charged-lepton and quark sectors. Thus even if all present anomalies evaporate, future precision measurements may uncover indirect NP effects that may, remarkably, be connected with the neutrino mass problem. Reference~\cite{Gargalionis:2020xvt} may be consulted for other interesting survey results, such as which pairs of exotics occur most commonly in these models.

\section{Closing remarks}

While we know that neutrinos have tiny but nonzero masses, we do not know what extended Lagrangian to write in the textbooks. There are many well motivated candidates, from high-scale to low-scale seesaw models, and radiative models. Obtaining the full answer is a distant goal, but we can reasonably hope for advances through the search for neutrinoless double beta decay, low mass heavy neutral leptons, charged-lepton flavour violation, deviations in $(g-2)_\ell$, NP quark flavour effects, and exotica production at colliders.

\acknowledgments This work was supported by the Australian Government, in part through the Australian Research Council Centre of Excellence for Dark Matter Particle Physics (CDM, CE200100008) and in part through the Australian Research Council Discovery Project DP200101470. I thank my students and collaborators for their hard work on joint papers relevant for this subject.

%\begin{thebibliography}{99}
%\bibitem{...}
%....

%\end{thebibliography}

\bibliographystyle{JHEP}
\bibliography{references}

\end{document}